\documentclass{article}
\usepackage[utf8]{inputenc}
\usepackage{fullpage}
\usepackage{graphicx}
\usepackage{tabularx}           
\usepackage{float}              
\usepackage[hidelinks]{hyperref}
\usepackage{setspace}
\usepackage{pgfplots}           
\usepackage{inconsolata}        
\usepackage{fnpct}              
\usepackage{microtype}          

\usetikzlibrary{external}

\pgfplotsset{compat=1.14}

\newcolumntype{b}{X}
\newcolumntype{s}{>{\hsize=.5\hsize}X}

\usepackage[inline]{enumitem}   
\usepackage{amsmath,amssymb}    

\title{Topological Descriptors Help Predict Guest Adsorption in Nanoporous Materials}

\author{Aditi S. Krishnapriyan\footnote{Stanford University, Stanford, California, United States of America} \and Maciej Haranczyk\footnote{IMDEA Materials Institute, C/Eric
Kandel 2, 28906 - Getafe, Madrid, Spain} \and Dmitriy Morozov\footnote{Lawrence Berkeley National Laboratory, Berkeley, California, United States of America}}
\date{}

\newcommand{\Vor}{\operatorname{Vor}}
\newcommand{\Alpha}{\operatorname{Alpha}}
\newcommand{\Rsp}{\mathbb{R}}
\newcommand{\Hgr}{{\sf H}}
\newcommand{\Dgm}{\operatorname{Dgm}}
\newcommand{\Gauss}{\mathcal{N}}

\begin{document}

\maketitle

\begin{abstract}
    Machine learning has emerged as an attractive alternative to experiments
    and simulations for predicting material properties. Usually, such an
    approach relies on specific domain knowledge for feature design: each
    learning target requires careful selection of features that an expert
    recognizes as important for the specific task.  The major drawback of this
    approach is that computation of only a few structural features has been
    implemented so far, and it is difficult to tell a priori which features are
    important for a particular application.
    The latter problem has been empirically observed for predictors of guest
    uptake in nanoporous materials: local and global porosity features become
    dominant descriptors at low and high pressures, respectively.

    We investigate a feature representation of materials using tools from
    topological data analysis.  Specifically, we use persistent homology to
    describe the geometry of nanoporous materials at various scales.
    We combine our topological descriptor with traditional
    structural features and investigate the relative importance of each to the
    prediction tasks.
    We demonstrate an application of this feature representation by predicting
    methane adsorption in zeolites, for pressures in the range of 1--200 bar.
    Our results not only show a considerable improvement compared to the
    baseline, but they also highlight that topological features capture
    information complementary to the structural features, which is especially
    important for the adsorption at low pressure, a task particularly difficult
    for the traditional features.
    %
    %
    Furthermore, by investigation of the importance of individual topological features in the adsorption model, we are able
    to pinpoint the location of the pores that correlate best to adsorption at
    different pressure, contributing to our atom-level understanding of structure-property relationships.
\end{abstract}



\section{Introduction}
\label{sec:intro}

Nanoporous materials, such as zeolites, are critical to many industrial sectors and are commonly used as membranes, adsorbents for separations, storage and delivery, and heterogeneous catalysts~\cite{phan_2010}. Along with zeolites, advanced porous materials (APMs) such as metal organic frameworks (MOFs), covalent organic frameworks (COFs), and related families of materials, have been investigated in the context of the above and other applications~\cite{zhou_2012}. An important factor that determines the utility of any nanoporous material is that its pore morphology and the corresponding material structures are configured to maximize the performance in its respective application. Finding the materials that satisfy the latter is therefore necessary to develop more efficient, cleaner, and cheaper technologies. The nanoporous materials space is immense: there are more than 1400 zeolite materials already reported that belong to more than 200 unique topologies~\cite{iza_zeolites}. However, these experimentally known zeolites constitute only a very small fraction of more than 2.7 million of hypothesised structures~\cite{pophale2011}, though the latter number may be orders of magnitude smaller than can be synthesized~\cite{salcedo_2019}. APMs have an even higher number of both already synthesized and predicted structures~\cite{yaghi2013,wilmer2012}, as their frameworks allow for greater flexibility in choosing diverse building blocks of various chemistries and topologies. 

Over the last decade, there has been a growing effort to employ computational techniques to navigate the space of possible nanoporous materials and choose promising candidates for various applications. Molecular simulations allow the study of important guest molecule phenomena in zeolites and APMs, such as adsorption and diffusion~\cite{snurr_2009}. Henry coefficients, heats of adsorption, adsorption isotherms, and diffusion coefficients of guest species can be predicted and there is solid evidence in the literature that these predictions are accurate~\cite{smit_2008,krishna_2006}. Initial attempts to screen for optimal materials have employed quite small collections of nanoporous materials, which could be characterized in their totality using molecular simulations (i.e. brute-force screening). The size of these studies were limited by the size of crystal structure databases, which at that time contained only experimentally known zeolites and early MOFs. The research landscape has changed dramatically with the introduction of databases of predicted nanoporous materials such as hypothetical zeolites~\cite{pophale2011}, which posed a significant computational challenge for the brute-force approach.

As a result, an entire spectrum of high-throughput screening tools and approaches have been developed in the last decade, which can be grouped into two categories. The first consists of techniques that limit the number of structures to be investigated using expensive molecular simulations, examples of which include 1) tools enabling screening filters based on porosity descriptors~\cite{sholl2010}, 2) substructure similarity screening and material diversity-based data reduction~\cite{martin2012,martin2012_2}, and 3) efficient searches using genetic algorithms or optimization approaches~\cite{chung2016}. The second category consists of techniques accelerating prediction of properties of each material in the considered set, examples of which are 1) GPU-accelerated molecular simulations~\cite{kim2012}, 2) simplified simulations and models, and finally 3) machine learning prediction trained using the simulation results of a relatively small subset of materials~\cite{bucior2018,simon2015}. The latter approach is of particular interest as there is sufficient evidence to show that it can provide high accuracy, robustness, and the throughput necessary to tackle ever-increasing datasets.


Machine learning models aim to predict the material property of choice by using a function fitted on a set of known properties and operating on a material representation as its input. There are a number of frameworks for these fittings that include algorithms such as random forests, neural networks, and support-vector machines, all of which have examples of successful applications in materials, and are often interchangeable. A suitable material representation poses a greater challenge. Typically, domain experts need to select
specific material features to serve as model input, used for making predictions about a
particular property. Standard examples of such features include density of
the material, cohesive energy per atom, and volume per atom~\cite{ward2018}.

In the context of nanoporous materials, one of the most desired properties to predict is a guest adsorption loading at given temperature--pressure conditions. Naturally, guest adsorption is determined by the morphology of the materials' pores, where guest molecules will be located and interacting with the material structure. Porosity descriptors such as accessible surface area (ASA), largest cavity diameter (LCD), and others have been primary targets to be used to construct feature vectors.
There are a growing number of models successfully trained and employed to predict adsorption of methane at conditions relevant to vehicular storage applications~\cite{sriva2017}, CO$_{2}$ in carbon capture~\cite{woo2014}, and Xe in Xe separations from air~\cite{simon2015}. These early studies have highlighted challenges related with the selection and implementation of relevant feature vectors.

In the case of high pressure gas adsorption, guest molecules generally occupy the entire void space available in a material. Therefore, inclusion of a ``global'' material porosity descriptor such as ASA leads to predictive models with these features becoming the most important to the model. Empirically, this and similar correlations have been observed: for example, the Chachine rule in hydrogen storage application, which states that one weight percent of H$_{2}$ can be stored per 500 m$^{2}$/g of a materials' ASA~\cite{chachine2011}.
However, in the case of low pressure adsorption, guest molecules tend to be localized in the strongly binding regions of the material's pore~\cite{lin2012}, which are typically not captured well with the standard porosity descriptors. To account for this, specific descriptors can be developed: for example, Voronoi energy was introduced~\cite{simon2015}, but at substantial development effort and with unknown application transferability potential.

The current state of the field suggests a significant need and potential payoff for the development of robust generic descriptors that would capture porosity features across pore scales and allow for generalized models for prediction of adsorption loadings at various pressures. 
In this paper, we focus on multi-scale topological features that describe the
connectivity of the pore space of the material.
Specifically, we use persistent
homology~\cite{edelsbrunner2007} to compute topological signatures of the
channels and voids constituting the porosity of a material, thereby capturing their number and size.
We convert these signatures to vector representations, facilitating 
supervised machine learning algorithms to predict material properties.
We note that persistent homology is very general and not tailored to a
specific application. Instead, the labels in the training data allow machine
learning algorithms to decide which subset of the features is important to a
particular problem.


While there has been previous work applying topological data analysis to
nanoporous materials~\cite{lee2018}, it focused on using standard, global
metrics on topological signatures to cluster the crystal structures.
A more recent work~\cite{Zhang2019} was also concerned with using topological
signatures for supervised learning of nanoporous material properties.  However,
the authors turned topological 
descriptors into simplified feature vectors,  which lost the ability to correlated the performance with specific topological features. 
In contrast, our goal is to use topological signatures with supervised machine learning to
predict a material property, while simultaneously identifying the parts of the
signature important for the prediction.

We demonstrate our approach in an application of prediction of methane adsorption in all silica zeolite structures, for which large sets of simulated data are available~\cite{simon2014}.
We show that our topological model outperforms existing porosity descriptors
when it comes to predicting adsorption properties, using a number of metrics
including root-mean-square error, R$^{2}$ score, and the Spearman correlation
coefficient. Furthermore, we show that unlike the handcrafted structural features,
the topological features give accurate predictions across a wide range of
pressures. We also show that the best model includes a combination of
topological and structural features, meaning that the two capture complementary
information about the material. 
Furthermore, our approach allows for identification of the structural features related with the important topological descriptors identified within the topological model, hence providing atom-level understanding of the origin of the property of interest.


\section{Methods}
\label{sec:methods}

\begin{figure}[h]
    \includegraphics[width=\textwidth]{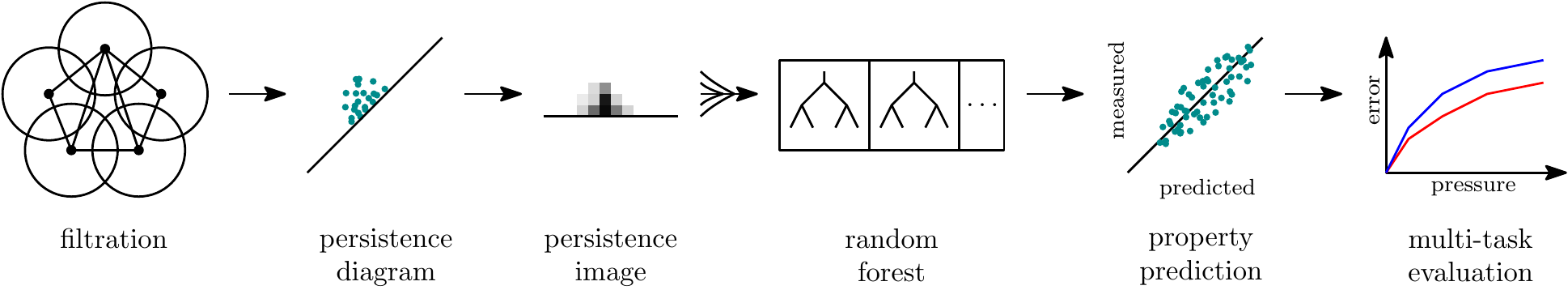}
    \caption{Computational pipeline.}
    \label{fig:pipeline}
\end{figure}

A schematic view of the computational pipeline is depicted in Figure~\ref{fig:pipeline}. In the initial step, material structures are analyzed to identify features present at various length scales. In particular, after normalizing the size of the structure representations, we compute their alpha shape filtrations,
whose persistence diagrams we use as the topological signature of each material.
Then, we vectorize these diagrams by turning them into so-called persistence images. Such vector representations are used together with the results of predicted guest uptakes  
to train random forest regression models for adsorption at different
pressures. Finally, we compare different measures of the quality of the regression across
choices of features and pressures. 
The rest of this section explains the methodology of the pipeline in more detail,
as well as the material dataset used to demonstrate the approach.

\paragraph{Zeolite morphology.}
Zeolites are crystalline structures. They are typically represented by the positions of
atoms with respect to the periodic cell, which can be 
implicitly replicated over an arbitrarily large volume. The cells
themselves have vastly different sizes. To normalize the calculation,
we prepare a 100\AA\ $\times$ 100\AA\ $\times$ 100\AA\ supercell of each
material. This size is chosen to be at least 2--3 times larger than the largest
base cell. 
A similar approach of representing materials by their spherical supercell was employed earlier~\cite{blastein}.

\begin{figure}
    \centering
    \includegraphics{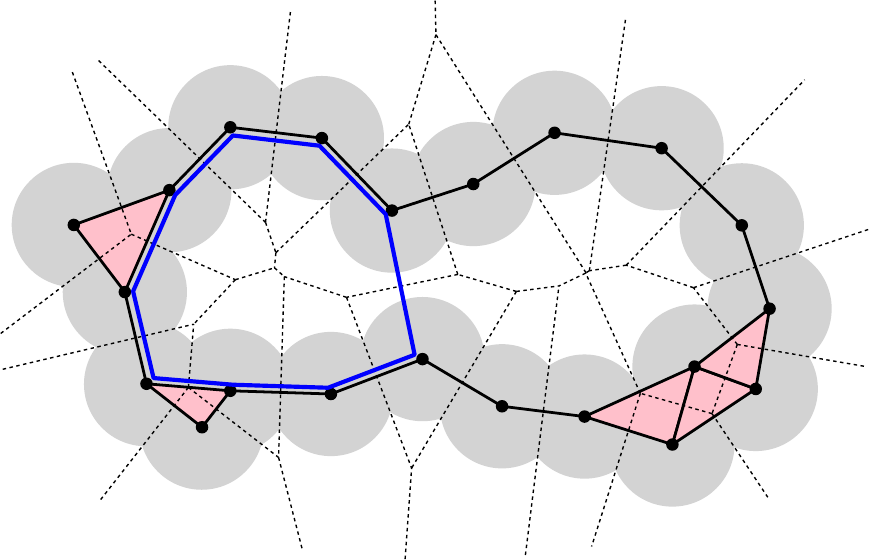}
    \hspace{.4in}
    \raisebox{0.2\height}{\includegraphics{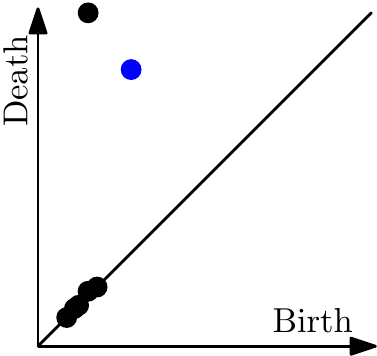}}
    \caption{Left: a point set $P$; a union of balls (gray) of fixed radius $r$,
             centered on the point set; a Voronoi diagram (dashed); an alpha
             shape $\Alpha_r(P)$ (points, solid edges, and pink triangles).
             Right: 1-dimensional persistence diagram of the alpha shape
             filtration of $P$; the blue point represents the cycle highlighted
             in blue on the left.}
    \label{fig:alpha-shapes}
\end{figure}

We model the pore structure of the materials across scales.
Specifically, we take the union of balls around the atoms and vary their radii.
We track how the topology of this union (and, dually, of its complement --- the
pore space) changes.
To represent this union combinatorially, we use alpha shapes. Briefly,
given a point set $P$, a \emph{Voronoi cell} of a point, $p$, consists of all points of
the ambient space that are closer to $p$ than to any other point in $P$:
\[
    \Vor(p) = \left\{ x \in \Rsp^3 \mid \|x - p\| \leq \| x - q \|\ \forall q \in P \right\}.
\]
An \emph{alpha shape} of $P$ for radius $r$ consists of all the subsets $\sigma$ of
$P$, where the intersection of the Voronoi cells and the balls of radius $r$
around the points $p \in \sigma$ is not empty:
\[
    \Alpha_r(P) =
        \left\{ \sigma \subset P \mid
                \bigcap_{p \in \sigma} B_r(p) \cap \Vor(p) \neq \emptyset \right\}.
\]
Figure~\ref{fig:alpha-shapes} illustrates a point set and its alpha shape for a
fixed radius.
Geometrically, an alpha shape consists of points, edges, triangles and
tetrahedra that are a subset of the Delaunay triangulation. More importantly for
us, an alpha shape for radius $r$ has the same topology as the union of balls of
the same radius.

By construction, alpha shapes nest as we increase their radius:
$\Alpha_r(P) \subseteq \Alpha_{r'}(P)$ for $r < r'$. Their nesting across all
radii forms an \emph{alpha shape filtration}.
We compute the alpha shapes using the software package
DioDe\footnote{\url{github.com/mrzv/diode}}, which is a Python binding
to the relevant functionality in the computational geometry library CGAL\footnote{\url{cgal.org}}.

\paragraph{Persistent homology.}
Persistent homology tracks the changes of homology groups in a filtration. We do
not define \emph{homology} formally --- there are thorough
surveys~\cite{edelsbrunner2007,EdMo17} of all the relevant concepts --- but note that this algebraic topological
invariant records the number of connected components, loops, and voids in a
space. Given an alpha shape filtration of a point set, the homology groups of
the nested spaces are connected by linear maps:
\[
    \Hgr(\Alpha_{r_1}(P)) \to
    \Hgr(\Alpha_{r_2}(P)) \to
    \ldots \to
    \Hgr(\Alpha_{r_n}(P)).
\]
By following the changes in the sequence, one can identify pairs of radii, when
a homology class (e.g., a loop or a void) appears and subsequently disappears:
i.e., when a tunnel or a pocket appears and then closes up in the
pore space. Such pairs of radii are called \emph{birth--death pairs}, $(b,d)$.
The bigger the difference, $d-b$, between death and birth, the more prominent,
or \emph{persistent}, is the class, and the more pronounced is the pore.

The full collection of birth--death pairs for a filtration is called a
persistence diagram, $\Dgm_k(P)$; see Figure~\ref{fig:alpha-shapes}. There are three non-trivial diagrams for the
alpha shapes in three dimensions. The 0-th diagram records the evolution of
connected components in the filtration; the 1-st diagram, loops; the 2-nd
diagram, voids. In the remainder, we use only the 1-st and 2-nd diagrams ---
the 0-th diagram, the structure of connected components, does not have much information beyond the
distances between nearest neighbor atoms.
We compute the persistence diagrams from alpha shapes using the software package
Dionysus\footnote{\url{https://github.com/mrzv/dionysus}}.

Persistent homology has many attractive properties. A crucial one has to do with
its resilience to noise: persistence is stable. If we perturb the point set a
little, the persistence diagram changes only a little. To be precise, let $d_H$
denote the Hausdorff distance between two points sets:
\[
    d_H(P,Q) = \max \left\{
                \max_{p \in P} \min_{q \in Q} \| p - q \|,
                \max_{q \in Q} \min_{p \in P} \| p - q \|
                    \right\}.
\]
Let $d_B$ denote the bottleneck distance between persistence diagrams:
\[
    d_B(\Dgm_k(P), \Dgm_k(Q)) =
        \inf_\gamma \sup_{x \in \Dgm_k(P)} \| x - \gamma(x) \|_\infty,
\]
where $\gamma: \Dgm_k(P) \to \Dgm_k(Q)$ is a bijection between the persistence
diagrams\footnote{For technical reasons, including to make the bijection $\gamma$
well-defined, persistence diagrams include all the points on the diagonal, taken
with infinite multiplicity. We do not dwell on this detail, but refer the
interested reader to the surveys~\cite{edelsbrunner2007,EdMo17}.}.
Stability means that the former upper bounds the latter:
\[
    d_B(\Dgm_k(P), \Dgm_k(Q)) \leq d_H(P,Q).
\]

\paragraph{Persistence images.}
Persistence diagrams are multi-sets of points in the plane. Most machine
learning algorithms operate on points in a fixed-dimensional Euclidean space. To
translate the former to the latter, Adams et al.~\cite{adams_jmlr_2017}
introduced \emph{persistence images}, which, after transforming the birth--death
pairs $(b,d)$ into birth--persistence pairs $(b,d-b)$, replace each one by a
Gaussian (to spread its influence across the neighborhood, since nearby points
represent features of similar size), and weigh each point in the plane:
\[
    f(x,y) = w(y) \sum_{(b,d) \in \Dgm(P)} \Gauss((b,d-b), \sigma).
\]
Persistence images result from discretizing $f(x,y)$ by integrating
its value within each grid cell of fixed dimensions.
Adams et al.\ choose the weighting so that $w(y) \to 0$ as $y \to 0$, for
example, the linear weighting, $w(y) = y$. This choice places more weight on
persistent points and ensures the stability of the persistence images. Since
we do not know a priori which pore size is significant for our application, and
because we are not as concerned with stability --- zeolite structures are
derived within crystal symmetry constraints and thus suffer at most from the
floating point error in the atomic coordinates --- we also experiment with using
no weighting, $w(y) = 1$.

We use a modified version of the PersIm
package\footnote{\url{persim.scikit-tda.org}} to compute persistence images.
We choose a resolution of $50 \times 50$ and a spread $\sigma = 0.15$.
We also tried spreads of 0.05, 0.10, 0.15, and 0.20, but the results had similar
accuracy in the range of 0.10--0.20. Similarly, we also tried different resolutions ranging from $10 \times 10$ to $100 \times 100$ and found that the middle range of $50 \times 50$ had the best accuracy. The images were scaled on the $x$-axis by the maximum birth
value across all diagrams and on the $y$-axis by the maximum persistence value
across all diagrams.


\paragraph{Data sets.}
To demonstrate an application of our approach, we have employed a dataset
constituting of crystal structures of all silica zeolites, for which methane
adsorption isotherms were determined previously~\cite{simon2014}.
Specifically, this set contains 55,039 structures composed on a subset of the PCOD
hypothetical zeolite database~\cite{pophale2011} and a subset of the
IZA-recognized zeolites~\cite{iza_zeolites}.
Adsorption isotherms were determined using grand-canonical Monte Carlo
simulations involving classical Lennard-Jones potential models the
framework--guest and guest--guest interactions~\cite{simon2014}. Isotherms were
sampled at 14 pressures in the 1--200 bar range.  

For each structure in our set, we have determined the values of the following commonly used descriptors which we refer to as baseline descriptors:
\begin{itemize}[nolistsep,noitemsep]
    \item
        Accessible volume (AV), in (cm\textsuperscript{3}/g), calculated as the total AV resulting from a summation of contributions from methane-accessible channels and non-accessible pockets;
    \item
        Accessible surface area (ASA), (cm\textsuperscript{3}/cm\textsuperscript{3}), also calculated with the inclusion of inaccessible pockets;
    \item
        Crystal density ($\rho$), in (kg/m\textsuperscript{3});
    \item
        Pore Limiting Diameter (PLD), in (\AA), the diameter of the largest sphere to percolate through material;
    \item
        Largest Cavity Diameter (LCD), in (\AA), the diameter of the largest sphere than can fit inside the material's pore system.
\end{itemize}
The values for the above descriptors 
were computed using the Zeo{\raisebox{.3ex}{\tiny\bf ++}} software package~\cite{zeoplusplus}.

\paragraph{Random forests.}
We use random forest~\cite{breiman_rf_2001}
regression to predict the methane adsorption at 14 different pressures, from 1 to 200 bar.
Random forests are ensembles of decision trees. Each decision tree is built
recursively by finding a feature, which allows for a cut that
maximizes the information gain about the target value.
Each tree in the ensemble is built on features chosen at random. The frequency
with which each feature is chosen for making a cut serves as an estimate of its
importance for the target problem. Such an ability to assign importance to
features and to combine features across different modalities, independently of
their scale, is the main reason for our choice of random forests.

We build the trees for different combinations of the features to evaluate their
relative importance. We use 1- and 2-dimensional persistence images separately and
together. Additionally, we employ baseline structural features, described in the previous paragraph,
as well as a combination of the two types of
features, topological and structural.

We train the random forest on the specific methane adsorption of each material
at a given pressure. Each of our forests consists of 500 trees, and the final
methane prediction is the average of the prediction of all the trees in the
forest. After training the random forest on the training set, predictions
are made on an unseen test set.
We evaluate the quality of the prediction by comparing the predicted and the
correct adsorption values. We compute the root-mean-square deviation,
$\sqrt{\sum (\hat{y}_i - y_i)^2 / n}$, the coefficient of determination (R$^2$),
$1 - \sum(y_i - \hat{y}_i)^2) / \sum (y_i - \bar{y})^2$,
and the Spearman rank coefficient, which computes how well the prediction
captures the ordered ranking of the materials by computing the Pearson
correlation between the ranks of the predicted and true methane adsorption
values, which represents a typical material screening application.
%


\section{Results}
\label{sec:results}


\begin{figure}[H]

\centering
\includegraphics[width=.99\textwidth]{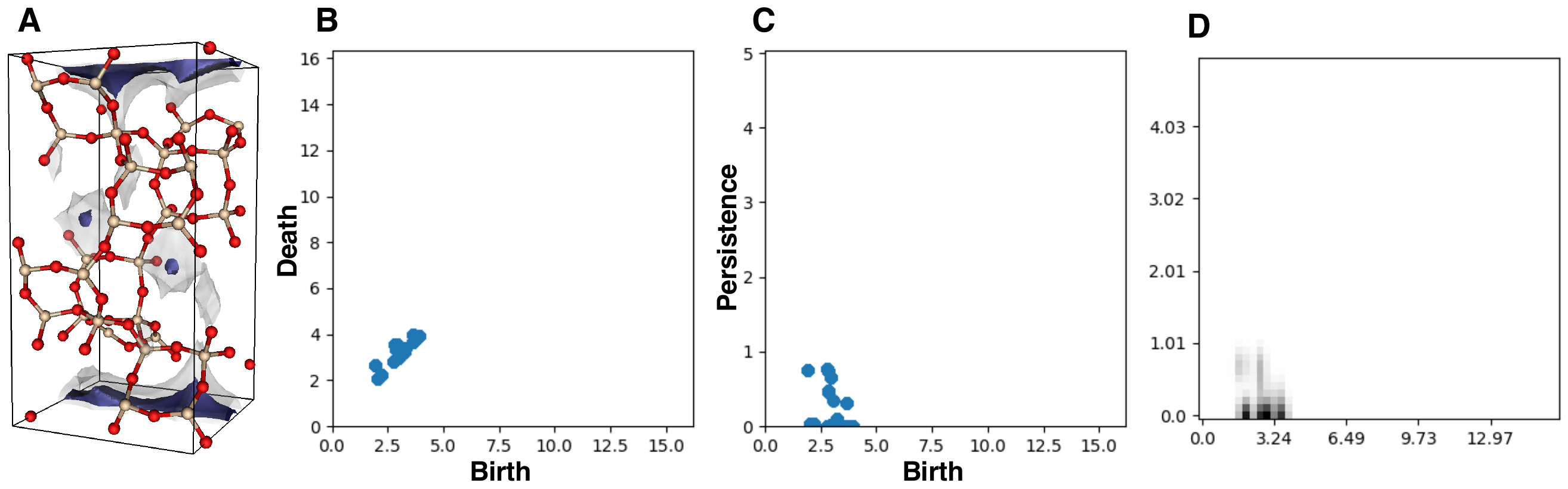}
\caption{Schematic outlining the (a) zeolite structure, (b) persistence diagram plotted as birth vs.\
         death, which is then (c) linearly transformed to birth vs.\ persistence, and
         then expressed as a (d) persistence image. All the diagrams are scaled based
         on the maximum birth, death, and persistence of the whole data set.}
\label{fig:pers_images}

\end{figure}


\pgfplotsset{
    cycle list={
    {red,  mark=none},     {red,  mark=none, dashed},
    {blue, mark=none},     {blue, mark=none, dashed},
    {green,  mark=none},   {green,  mark=none, dashed},
    {brown, mark=none,   very thick},  {brown, mark=none, dashed, very thick},
    {magenta, mark=none, very thick},  {magenta, mark=none, dashed},
    }
}

\begin{figure}[H]
\begin{minipage}{0.5\textwidth}
\centering
\begin{tikzpicture}
\begin{axis}[
                symbolic x coords={1.0-bar, 5.8-bar, 10.0-bar, 15.0-bar,
                                   20.0-bar, 25.0-bar, 30.0-bar, 35.0-bar, 40.0-bar, 45.0-bar,
                                   50.0-bar, 65.0-bar, 100.0-bar, 200.0-bar},
                xtick={1.0-bar, 10.0-bar, 20.0-bar, 30.0-bar, 40.0-bar, 50.0-bar, 200.0-bar},
                xticklabels={1, 10, 20, 30, 40, 50, 200},
                height=3in,
                width=.99\textwidth,
                every axis plot/.append style={thick},
                ylabel={RMSD},
                legend columns=1,
                legend to name=rmsd-legend,
                legend style={draw=none},
                legend cell align={left},
            ]
    \addplot    +[] table[x=pressure, y=topology-1d-linear-weighting]                   {data/rmsd_actual_transposed.txt}; \addlegendentry{1D topology, linear weighting}
    \addplot    +[] table[x=pressure, y=topology-1d-no-weighting]                       {data/rmsd_actual_transposed.txt}; \addlegendentry{1D topology, no weighting}
    \addplot    +[] table[x=pressure, y=topology-2d-linear-weighting]                   {data/rmsd_actual_transposed.txt}; \addlegendentry{2D topology, linear weighting}
    \addplot    +[] table[x=pressure, y=topology-2d-no-weighting]                       {data/rmsd_actual_transposed.txt}; \addlegendentry{2D topology, no weighting}
    \addplot    +[] table[x=pressure, y=topology-linear-weighting]                      {data/rmsd_actual_transposed.txt}; \addlegendentry{Total topology, linear weighting}
    \addplot    +[] table[x=pressure, y=topology-no-weighting]                          {data/rmsd_actual_transposed.txt}; \addlegendentry{Total topology, no weighting}
    \addplot    +[] table[x=pressure, y=baseline-topology-combined-linear-weighting]   {data/rmsd_actual_transposed.txt}; \addlegendentry{Combined, linear weighting}
    \addplot    +[] table[x=pressure, y=baseline-topology-combined-no-weighting]       {data/rmsd_actual_transposed.txt}; \addlegendentry{Combined, no weighting}
    \addplot    +[] table[x=pressure, y=baseline]                                      {data/rmsd_actual_transposed.txt}; \addlegendentry{Baseline}
\end{axis}
\end{tikzpicture}
\end{minipage}
\hfill
\begin{minipage}{0.4\textwidth}
    \ref{rmsd-legend}
\end{minipage}

\begin{minipage}{0.5\textwidth}
\centering
\begin{tikzpicture}
\begin{axis}[
                symbolic x coords={1.0-bar, 5.8-bar, 10.0-bar, 15.0-bar,
                                   20.0-bar, 25.0-bar, 30.0-bar, 35.0-bar, 40.0-bar, 45.0-bar,
                                   50.0-bar, 65.0-bar, 100.0-bar, 200.0-bar},
                xtick={1.0-bar, 10.0-bar, 20.0-bar, 30.0-bar, 40.0-bar, 50.0-bar, 200.0-bar},
                xticklabels={1, 10, 20, 30, 40, 50, 200},
                height=3in,
                width=.99\textwidth,
                xlabel={Pressure (bar)},
                every axis plot/.append style={thick},
                ylabel={R$^2$},
                legend columns=1,
                legend to name=r2-legend,
                legend style={draw=none},
                legend cell align={left},
            ]
    \addplot    +[] table[x=pressure, y=topology-1d-linear-weighting]                   {data/r2.txt}; \addlegendentry{1D topology, linear weighting}
    \addplot    +[] table[x=pressure, y=topology-1d-no-weighting]                       {data/r2.txt}; \addlegendentry{1D topology, no weighting}
    \addplot    +[] table[x=pressure, y=topology-2d-linear-weighting]                   {data/r2.txt}; \addlegendentry{2D topology, linear weighting}
    \addplot    +[] table[x=pressure, y=topology-2d-no-weighting]                       {data/r2.txt}; \addlegendentry{2D topology, no weighting}
    \addplot    +[] table[x=pressure, y=topology-linear-weighting]                      {data/r2.txt}; \addlegendentry{Total topology, linear weighting}
    \addplot    +[] table[x=pressure, y=topology-no-weighting]                          {data/r2.txt}; \addlegendentry{Total topology, no weighting}
    \addplot    +[] table[x=pressure, y=baseline-topology-combined-linear-weighting]    {data/r2.txt}; \addlegendentry{Combined, linear weighting}
    \addplot    +[] table[x=pressure, y=baseline-topology-combined-no-weighting]        {data/r2.txt}; \addlegendentry{Combined, no weighting}
    \addplot    +[] table[x=pressure, y=baseline]                                       {data/r2.txt}; \addlegendentry{Baseline}
\end{axis}
\end{tikzpicture}
\end{minipage}
\hfill
\begin{minipage}{0.5\textwidth}
\centering
\begin{tikzpicture}
\begin{axis}[
                symbolic x coords={1.0-bar, 5.8-bar, 10.0-bar, 15.0-bar,
                                   20.0-bar, 25.0-bar, 30.0-bar, 35.0-bar, 40.0-bar, 45.0-bar,
                                   50.0-bar, 65.0-bar, 100.0-bar, 200.0-bar},
                xtick={1.0-bar, 10.0-bar, 20.0-bar, 30.0-bar, 40.0-bar, 50.0-bar, 200.0-bar},
                xticklabels={1, 10, 20, 30, 40, 50, 200},
                height=3in,
                width=.99\textwidth,
                xlabel={Pressure (bar)},
                every axis plot/.append style={thick},
                ylabel={Spearman correlation},
                ylabel near ticks,
                yticklabel pos=right,
                legend columns=1,
                legend to name=spearman-legend,
                legend style={draw=none},
                legend cell align={left},
            ]
    \addplot    +[] table[x=pressure, y=topology-1d-linear-weighting]                   {data/spearman.txt}; \addlegendentry{1D topology, linear weighting}
    \addplot    +[] table[x=pressure, y=topology-1d-no-weighting]                       {data/spearman.txt}; \addlegendentry{1D topology, no weighting}
    \addplot    +[] table[x=pressure, y=topology-2d-linear-weighting]                   {data/spearman.txt}; \addlegendentry{2D topology, linear weighting}
    \addplot    +[] table[x=pressure, y=topology-2d-no-weighting]                       {data/spearman.txt}; \addlegendentry{2D topology, no weighting}
    \addplot    +[] table[x=pressure, y=topology-linear-weighting]                      {data/spearman.txt}; \addlegendentry{Total topology, linear weighting}
    \addplot    +[] table[x=pressure, y=topology-no-weighting]                          {data/spearman.txt}; \addlegendentry{Total topology, no weighting}
    \addplot    +[] table[x=pressure, y=baseline-topology-combined-linear-weighting]    {data/spearman.txt}; \addlegendentry{Combined, linear weighting}
    \addplot    +[] table[x=pressure, y=baseline-topology-combined-no-weighting]        {data/spearman.txt}; \addlegendentry{Combined, no weighting}
    \addplot    +[] table[x=pressure, y=baseline]                                       {data/spearman.txt}; \addlegendentry{Baseline}
\end{axis}
\end{tikzpicture}
\end{minipage}
\caption{Comparison of root-mean-square deviation (top-left),
         coefficient of determination (bottom-left), and
         Spearman rank coefficient (bottom-right) in predicting
         methane adsorption for different features at different pressures.
         Top-right: plot legend.}
\label{fig:metrics}
\end{figure}
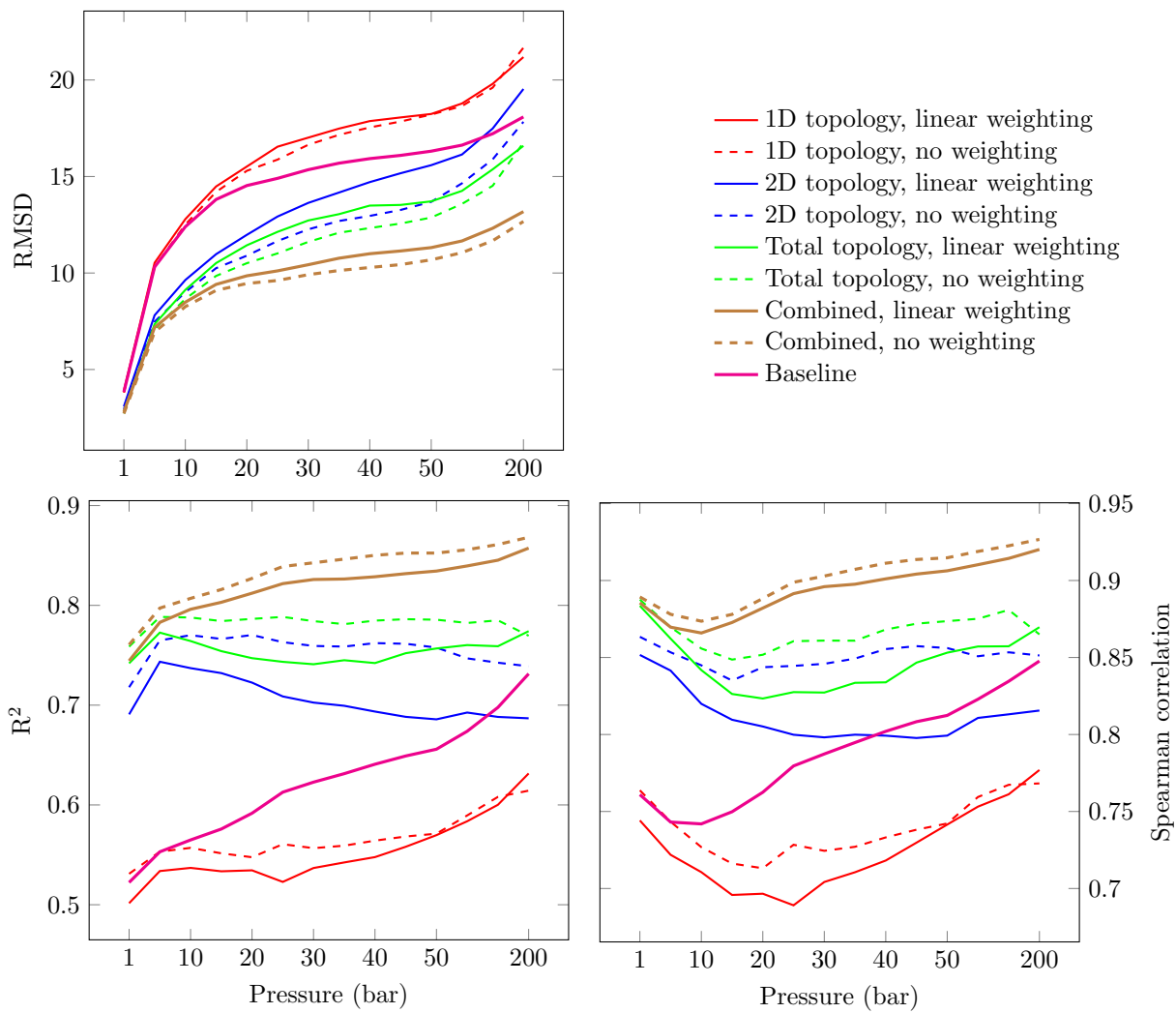

Figure~\ref{fig:metrics} shows the performance of different sets of features in
predicting methane adsorption in the 1-200 bar pressure range, measured using the root-mean-square deviation
(RMSD), the R$^{2}$ score, and the Spearman rank coefficient. The RMSD is low at lower pressures because the distribution of methane adsorption has low variance
in that regime.  The overall trend of all three metrics is the same:
the baseline features perform better than 1-dimensional topological features
alone; they also perform better than 2-dimensional features with linear weights
at higher pressures.  2-dimensional topological features perform better than
baseline (except with linear weights at higher pressures), and combined
1- and 2-dimensional topological features always perform better than the
baseline. However, the best results are derived from combining topological and
baseline structural features. This clearly shows that topology contributes
complimentary information to the baseline and significantly improves performance
of our regression model (with R$^2$ increasing from .52 to .76 at 1 bar and from .73
to .87 at 200 bar).  





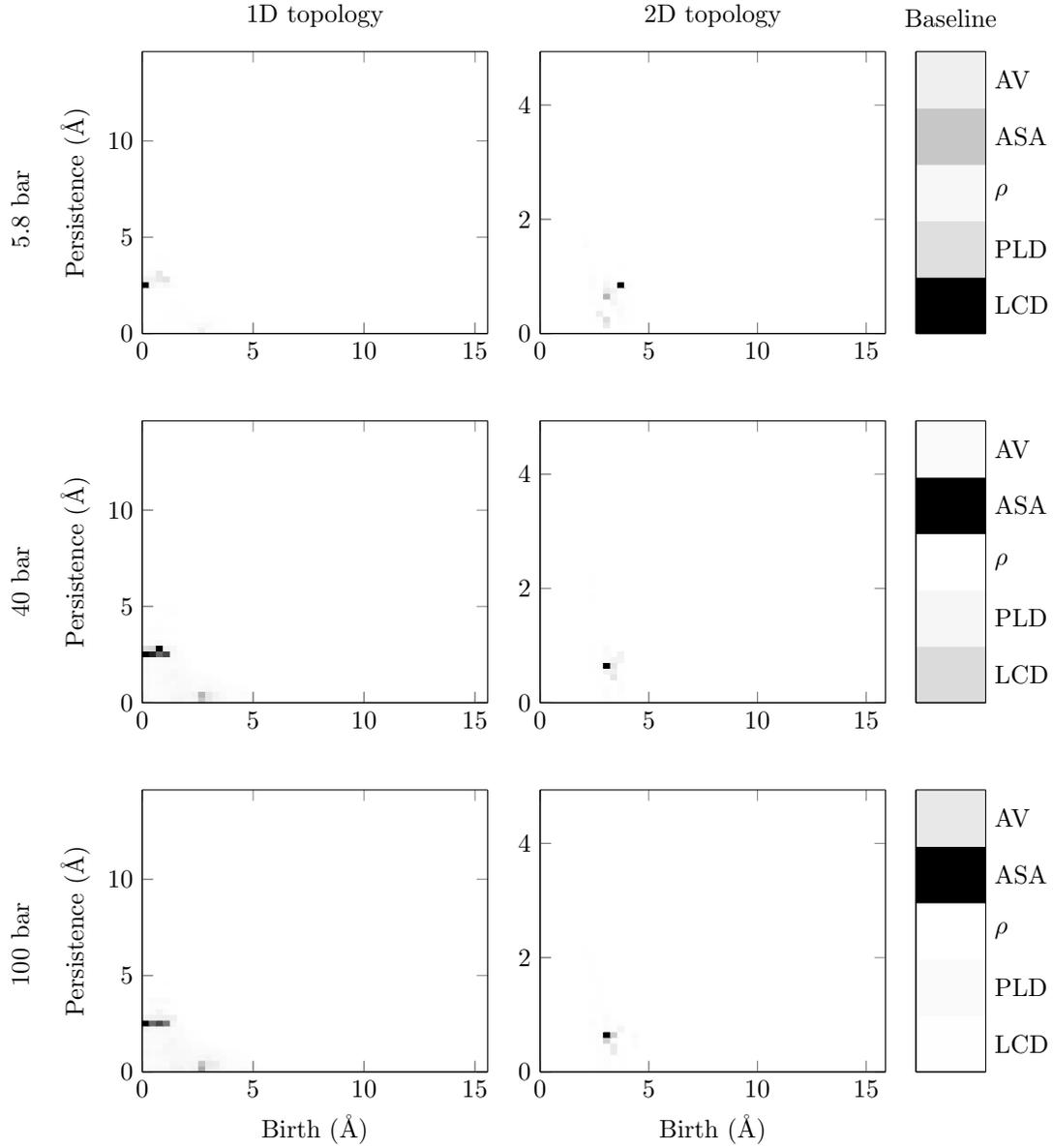
\begin{figure}[H]
\centering

\begin{minipage}[t]{3ex}
    \rotatebox{90}{\hspace{7ex} 5.8 bar}
\end{minipage}
\begin{minipage}[t]{0.3\textwidth}
\begin{tikzpicture}[baseline]
\begin{axis}[
        view={0}{90},
        height=\axisdefaultheight*.75,
        ylabel=Persistence (\AA),
        colormap = {whiteblack}{color(0cm)  = (white);color(1cm) = (black)},
        x coord trafo/.code={\pgfmathparse{(#1)/50*15.889533042907715}},
        y coord trafo/.code={\pgfmathparse{(#1)/50*14.93587589263916}},
        title=1D topology
    ]
    \addplot3[surf, shader=flat corner] file {data/1d-feature-importance-5.8.txt};
\end{axis}
\end{tikzpicture}
\end{minipage}
\hspace{3ex}
\begin{minipage}[t]{0.3\textwidth}
\begin{tikzpicture}[baseline]
\begin{axis}[
        view={0}{90},
        height=\axisdefaultheight*.75,
        ylabel=\phantom{Persistence (\AA)},
        colormap = {whiteblack}{color(0cm)  = (white);color(1cm) = (black)},
        x coord trafo/.code={\pgfmathparse{(#1)/50*16.215728759765625}},
        y coord trafo/.code={\pgfmathparse{(#1)/50*5.035214900970459}},
        title=2D topology
    ]
    \addplot3[surf, shader=flat corner] file {data/2d-feature-importance-5.8.txt};
\end{axis}
\end{tikzpicture}
\end{minipage}
\hspace{5ex}
\begin{minipage}[t]{0.2\textwidth}
\begin{tikzpicture}[baseline]
\begin{axis}[
        view={0}{90},
        height=\axisdefaultheight*.75,
        width=1in,
        colormap = {whiteblack}{color(0cm)  = (white);color(1cm) = (black)},
        ytick={0.5,1.5,2.5,3.5,4.5},
        yticklabels={LCD,PLD,$\rho$,ASA, AV},
        yticklabel pos=right,
        ytick style={draw=none},
        xtick=\empty,
        title=Baseline
    ]
    \addplot3[surf,shader=flat corner] file {data/baseline-feature-importance-5.8.txt};
\end{axis}
\end{tikzpicture}
\end{minipage}

\begin{minipage}[t]{3ex}
    \rotatebox{90}{\hspace{7ex} 40 bar}
\end{minipage}
\begin{minipage}[t]{0.3\textwidth}
\begin{tikzpicture}[baseline]
\begin{axis}[
        view={0}{90},
        height=\axisdefaultheight*.75,
        ylabel=Persistence (\AA),
        colormap = {whiteblack}{color(0cm)  = (white);color(1cm) = (black)},
        x coord trafo/.code={\pgfmathparse{(#1)/50*15.889533042907715}},
        y coord trafo/.code={\pgfmathparse{(#1)/50*14.93587589263916}},
    ]
    \addplot3[surf, shader=flat corner] file {data/1d-feature-importance-40.txt};
\end{axis}
\end{tikzpicture}
\end{minipage}
\hspace{3ex}
\begin{minipage}[t]{0.3\textwidth}
\begin{tikzpicture}[baseline]
\begin{axis}[
        view={0}{90},
        height=\axisdefaultheight*.75,
        ylabel=\phantom{Persistence (\AA)},
        colormap = {whiteblack}{color(0cm)  = (white);color(1cm) = (black)},
        x coord trafo/.code={\pgfmathparse{(#1)/50*16.215728759765625}},
        y coord trafo/.code={\pgfmathparse{(#1)/50*5.035214900970459}},
    ]
    \addplot3[surf, shader=flat corner] file {data/2d-feature-importance-40.txt};
\end{axis}
\end{tikzpicture}
\end{minipage}
\hspace{5ex}
\begin{minipage}[t]{0.2\textwidth}
\begin{tikzpicture}[baseline]
\begin{axis}[
        view={0}{90},
        height=\axisdefaultheight*.75,
        width=1in,
        colormap = {whiteblack}{color(0cm)  = (white);color(1cm) = (black)},
        ytick={0.5,1.5,2.5,3.5,4.5},
        yticklabels={LCD,PLD,$\rho$,ASA, AV},
        yticklabel pos=right,
        ytick style={draw=none},
        xtick=\empty,
        title=\phantom{Baseline}
    ]
    \addplot3[surf,shader=flat corner] file {data/baseline-feature-importance-40.txt};
\end{axis}
\end{tikzpicture}
\end{minipage}

\begin{minipage}[t]{3ex}
    \rotatebox{90}{\hspace{7ex} 100 bar}
\end{minipage}
\begin{minipage}[t]{0.3\textwidth}
\begin{tikzpicture}[baseline]
\begin{axis}[
        view={0}{90},
        height=\axisdefaultheight*.75,
        xlabel=Birth (\AA),
        ylabel=Persistence (\AA),
        colormap = {whiteblack}{color(0cm)  = (white);color(1cm) = (black)},
        x coord trafo/.code={\pgfmathparse{(#1)/50*15.889533042907715}},
        y coord trafo/.code={\pgfmathparse{(#1)/50*14.93587589263916}},
    ]
    \addplot3[surf, shader=flat corner] file {data/1d-feature-importance-100.txt};
\end{axis}
\end{tikzpicture}
\end{minipage}
\hspace{3ex}
\begin{minipage}[t]{0.3\textwidth}
\begin{tikzpicture}[baseline]
\begin{axis}[
        view={0}{90},
        height=\axisdefaultheight*.75,
        xlabel=Birth (\AA),
        ylabel=\phantom{Persistence (\AA)},
        colormap = {whiteblack}{color(0cm)  = (white);color(1cm) = (black)},
        x coord trafo/.code={\pgfmathparse{(#1)/50*16.215728759765625}},
        y coord trafo/.code={\pgfmathparse{(#1)/50*5.035214900970459}},
    ]
    \addplot3[surf, shader=flat corner] file {data/2d-feature-importance-100.txt};
\end{axis}
\end{tikzpicture}
\end{minipage}
\hspace{5ex}
\begin{minipage}[t]{0.2\textwidth}
\begin{tikzpicture}[baseline]
\begin{axis}[
        view={0}{90},
        height=\axisdefaultheight*.75,
        width=1in,
        colormap = {whiteblack}{color(0cm)  = (white);color(1cm) = (black)},
        ytick={0.5,1.5,2.5,3.5,4.5},
        yticklabels={LCD,PLD,$\rho$,ASA,AV},
        yticklabel pos=right,
        ytick style={draw=none},
        xtick=\empty,
        title=\phantom{Baseline}
    ]
    \addplot3[surf,shader=flat corner] file {data/baseline-feature-importance-100.txt};
\end{axis}
\end{tikzpicture}
\end{minipage}

\caption{Feature importance of the 1D topology, 2D topology, and baseline
         features at low (5.8 bar), medium (40 bar), and high (100 bar) pressures.
         The feature importance colormap is scaled relative to each individual
         plot. The relative contribution of 1D, 2D topology and baseline
         features appears in Figure~\ref{fig:feature-importance-summary}.}
\label{fig:feature-importance}
\end{figure}

A particularly interesting revelation of our experiments is the role the
weights, used to construct the persistence images, play in the performance of
our models.  The linear weights, which guarantee stability of the persistence
images, perform worse for all but the highest pressure than no weights, which
make no such guarantees. Our best explanation of this behavior is the following.
Linear weights make an implicit judgement about the significance of the width of
the channels and the diameter of the voids, determining the narrower and smaller
topological features (those with lower persistence) to be less important. This
scheme is reasonable for many uses of persistence, and indeed ensures the
stability of persistence images by limiting the effect of the smaller
topological features that result from noisy perturbation of the data. However,
with zeolite structures, there is no such error. The atomic coordinates lie on a
crystal lattice and the only error we expect is the floating point error. On the other
hand, small topological features are related to the roughness of the pore
surface, making them important to our application. These observations highlight
a major strength of our approach: instead of relying on the pre-determined
weights that assign significance to the different regions of the persistence
diagrams and images, we let the supervision involved in the learning process to
assign the appropriate weights to the different parts of the images. As
Figure~\ref{fig:feature-importance} illustrates, the weights the algorithm
chooses do not decay with persistence. 

The ability to determine the weights --- more accurately, the
feature importance --- in our model is one of our main motivations for using the
random forest. This algorithm builds an ensemble of decision trees, choosing a
random subset of features for each one. The frequency with which a particular
feature is chosen for a split, across the ensemble, is its
importance.  Figure~\ref{fig:feature-importance} shows the importance of the
features for the 1-, 2-dimensional topological and the baseline features at
three different pressures (low, medium, and high).
The color map differs for each plot (even within a single pressure), to make
the low values more visible. 

\begin{figure}[H]

\centering
\includegraphics[width=.4\textwidth]{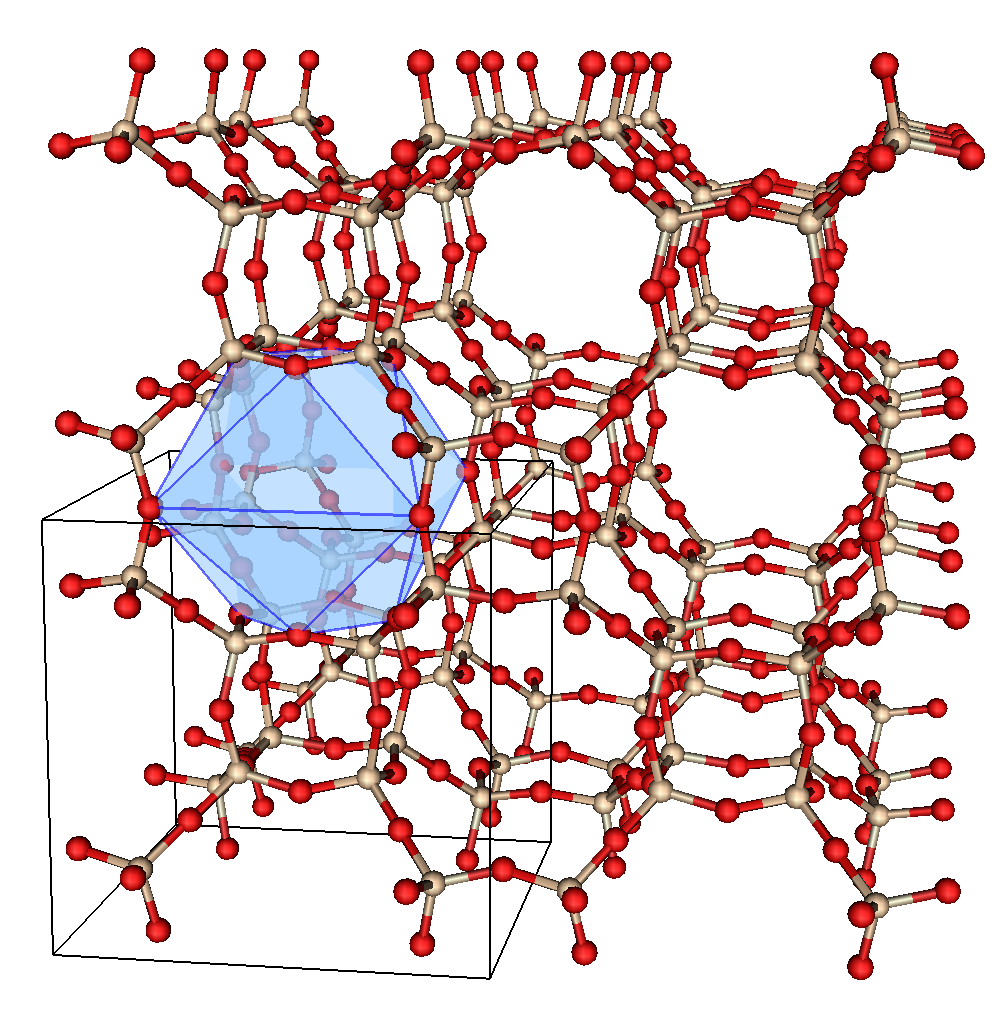}
\caption{Representative cycle (as highlighted by the points enclosing the blue structure) of the most important 2-dimensional topology features of a high methane adsorption zeolite, ACO. The porous structure corresponds to the birth and persistence values in Figure~\ref{fig:feature-importance} that are most predictive of methane adsorption.}
\label{fig:rep_cycle}

\end{figure}

From these plots, it is possible to pinpoint the birth and persistence values that correlate most strongly to predicting methane adsorption. As these values describe the voids in a crystal structure, they directly correspond to the structure of the zeolite and thus make it possible to elucidate the porous structure of materials with high methane adsorption. Figure~\ref{fig:rep_cycle} shows an example of the representative cycle for the most important 2-dimensional topology features for a zeolite (ACO) with high methane adsorption, i.e. the porous structure corresponding to the birth and persistence values in Figure~\ref{fig:feature-importance}. A notable property of nanoporous materials is their tunability, i.e. the ability to produce a material with precisely sized pores. Thus, understanding the porous structure correlating to a material property of interest provides insight into synthesizing such a material.

\begin{figure}[H]
\centering
\begin{minipage}{0.5\textwidth}
\centering
\begin{tikzpicture}[baseline]
\begin{axis}[
                title={Relative feature importance},
                ybar stacked,
                symbolic x coords={1.0-bar, 5.8-bar, 10.0-bar, 15.0-bar,
                                   20.0-bar, 25.0-bar, 30.0-bar, 35.0-bar, 40.0-bar, 45.0-bar,
                                   50.0-bar, 65.0-bar, 100.0-bar, 200.0-bar},
                xtick={1.0-bar, 10.0-bar, 20.0-bar, 30.0-bar, 40.0-bar, 50.0-bar, 200.0-bar},
                xticklabels={1, 10, 20, 30, 40, 50, 200},
                height=2.5in,
                width=.99\textwidth,
                xlabel={Pressure (bar)},
                reverse legend,
                legend style={draw=none},
                legend columns=1,
                legend style={at={(1.4,.6)}},
                legend cell align={left},
            ]
    \addplot    +[black,fill=magenta!60!white] table[x=pressure, y=baseline]                   {data/feature_summations.txt}; \addlegendentry{Baseline}
    \addplot    +[black,fill=blue!60!white]    table[x=pressure, y=2d-topology]                {data/feature_summations.txt}; \addlegendentry{2D topology}
    \addplot    +[black,fill=red!60!white]     table[x=pressure, y=1d-topology]                {data/feature_summations.txt}; \addlegendentry{1D topology}
\end{axis}
\end{tikzpicture}
\end{minipage}
\caption{Summary of relative feature importance across different pressures for
         the 1D, 2D topological features, and baseline features. Combined
         topological features dominate for low pressures, but as pressure
         increases, the baseline features become more important.  2D topology
         features remain significant even at higher pressures.}
\label{fig:feature-importance-summary}
\end{figure}
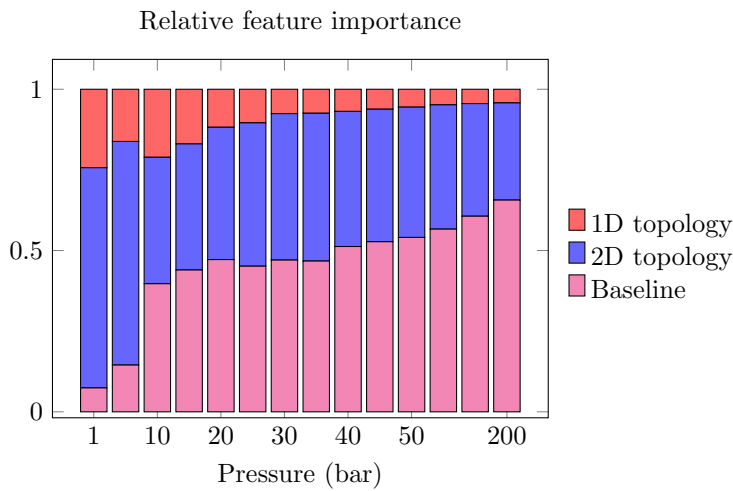

The relative importance of the three different
types of features is shown in Figure~\ref{fig:feature-importance-summary}.
The 1- and 2-dimensional topology plots show the regions of birth and persistence
values, which are the best predictors for methane adsorption. 
Although they change across pressures, these changes are small.
In contrast, LCD is the most important feature at low pressures, but ASA
becomes the most important feature at higher pressures. It numerically illustrates the empirical observations from earlier studies that local and global features are less transferable
across different pressures in comparison to the topology-based features.

Figure~\ref{fig:feature-importance-summary} shows how the relative importance of
the three different types of features changes with pressure. Each bar is the sum
of feature importance in the combined model, using all three types of features.
The plot confirms what we have already observed in Figure~\ref{fig:metrics}. At
low pressures, topological features perform significantly better than the
baseline features and the model places most of its weight on them. As the
pressure increases, more weight is assigned to the baseline features, although
even at the highest pressure, the combined weight on the topological features
exceeds $34\%$ (with most weight assigned to the 2D topological features).

A more recent work~\cite{Zhang2019} is also concerned with using topological
signatures for supervised learning of nanoporous material properties.  However,
there is a key difference from our work: the way it turns topological
descriptors into feature vectors relies on constructing ad hoc handcrafted
summaries, making it impossible to localize its important subsets.
Accordingly, as the authors show~\cite{Zhang2019}, the most important of these
(global) features closely correlate with the standard structural features,
namely, ASA and LCD in our terminology above.
In contrast, we input topological descriptors into machine learning algorithms
with minimal processing, allowing us both to extract finer information about
their important subsets and to show that they capture information complementary
to the traditional structural features.
Although Zhang et al.~\cite{Zhang2019} consider a slightly different learning task than
us, it is sufficiently close for a reasonable comparison: we report higher
R$^2$ scores.

\section{Conclusions}

We have developed an approach using topology to create a more holistic
feature representation for nanoporous materials than the currently widely used porosity descriptors. 
Specifically, this representation is a vectorized persistence diagram obtained from analysis of the normalized supercell representation of the materials' crystal structure, that can then be utilized in any machine learning algorithm. 

We have tested this approached using a set of roughly 55 thousand zeolite structures and their methane adsorption isotherms obtained using molecular simulations, which enable us to train random forest regressors operating on our representations. We have demonstrated that this
approach performs better than current scientific structural features and can accurately predict both adsorption loadings and structures rankings, highlighting the possibility of its applications in various screening studies.

We also note that the highest performance of statistical models was achieved by combining the new topology representations with the commonly used porosity descriptors.

\section*{Acknowledgements}

This work was partially supported by Laboratory Directed Research and Development (LDRD)
funding from Berkeley Lab, provided by the Director, Office of Science, of the
U.S. Department of Energy under Contract No. DE-AC02-05CH11231.
This research used resources of the National Energy Research Scientific
Computing Center (NERSC), a U.S. Department of Energy Office of Science User
Facility operated under Contract No. DE-AC02-05CH11231.
M.H. acknowledges support from the Spanish Ministry of Economy and Competitiveness
(RYC-2013-13949). 

\bibliographystyle{unsrt}
\bibliography{main}

\end{document}